\documentclass[prb,twocolumn,showpacs,amsmath,amssymb,superscriptaddress]{revtex4-1}

\usepackage{graphicx}
\usepackage{bm}
\usepackage{multirow}
\usepackage{hyperref}
\usepackage{bbold}
\hypersetup{backref=true,
 pdfnewwindow=true, colorlinks=true,
 linkcolor=blue, anchorcolor=blue,
 citecolor=blue, filecolor=blue,
 menucolor=blue, urlcolor=blue}

\def\kA{{\color{blue}$\downarrow$}\ }
\def\kB{{\color{red}$\uparrow$}\ }

\begin{document}

\title{Alternative structure of TiO$_2$ with higher energy valence band edge}

\author{Sinisa Coh}
\altaffiliation[Current address: ]{Mechanical Engineering, Materials Science and Engineering, University of California Riverside, Riverside, CA 92521, USA}
\email{sinisacoh@gmail.com} 
\affiliation{Department of Physics, University of California, and
  Materials Sciences Division, Lawrence Berkeley National Laboratory,
  Berkeley, CA 94720, USA}
\author{Peter Y. Yu}
\affiliation{Department of Physics, University of California, and
  Materials Sciences Division, Lawrence Berkeley National Laboratory,
  Berkeley, CA 94720, USA}
\author{Yuta Aoki}
\affiliation{Department of Physics, Tokyo Institute of Technology,
  2-12-1 Oh-okayama, Meguro-ku, Tokyo 152-8551, Japan}
\author{Susumu Saito}
\affiliation{Department of Physics, Tokyo Institute of Technology,
  2-12-1 Oh-okayama, Meguro-ku, Tokyo 152-8551, Japan}
\author{Steven G. Louie} 
\affiliation{Department of Physics, University of California, and
  Materials Sciences Division, Lawrence Berkeley National Laboratory,
  Berkeley, CA 94720, USA}
\author{Marvin L. Cohen} 
\affiliation{Department of Physics, University of California, and
  Materials Sciences Division, Lawrence Berkeley National Laboratory,
  Berkeley, CA 94720, USA}

\date{\today}

\pacs{71.15.Nc,61.46.Hk}

\begin{abstract}
We propose an alternative structure of TiO$_2$ anatase that has a higher energy oxygen $p$-like valence band maximum than the pristine TiO$_2$ anatase and thus has a much better alignment with the water splitting levels.  This alternative structure is unique when considering a large subspace of possible structural distortions of TiO$_2$ anatase.  We propose two routes towards this state and argue that one of them might have been realized in the recently discovered so-called black TiO$_2$.
\end{abstract}

\maketitle

\section{Background}
\label{sec:introduction}

Finding a suitable photocatalytic material that enables sunlight to split water into hydrogen and oxygen  has the potential to impact one of the major societal challenges today---clean production of energy.  Titanium dioxide (TiO$_2$) is nearly an ideal photocatalyst\cite{Fujishima1972} for this purpose because it is inexpensive and chemically stable.  Unfortunately the electronic band gap of TiO$_2$ is so large\cite{PhysRevB.18.5606} ($>$3~eV) that it absorbs only $\sim$5\% of the solar spectrum.  Therefore a large effort has been made to band-engineer TiO$_2$ so that it can harvest a larger portion of the solar spectrum.  As is well established,\cite{Xu543} the oxygen $p$-like valence band of TiO$_2$ is lower than the water-splitting level by $\sim$2~eV, while the conduction band is well aligned with the water-splitting level.  Therefore, it seems that the simplest strategy to reduce the band gap of TiO$_2$ --- without affecting its photocatalytic properties --- is to move its valence band to a higher energy and leave its conduction band energy intact. 

Doping is a common strategy to band-engineer TiO$_2$.\cite{Asahi269, Khan2243, Devi2013559}  However, doping is known to reduce photocatalytic efficiency since it introduces recombination centers.\cite{doi:10.1021/j100102a038}

Recently a new form of disordered TiO$_2$, so-called black TiO$_2$, was fabricated by hydrogenating pure, dopant-free, anatase-TiO$_2$ nano-sized crystals under pressure.\cite{Chen2011} Hydrogenation reduced the band gap of anatase TiO$_2$ from 3.3~eV to 1.5~eV which is a much better matched to the solar spectrum.  Most importantly, there are experimental indications that this gap reduction was formed by a movement of the valence band while the conduction band was mostly unaffected by the hydrogenation process.  Therefore, black TiO$_2$ has a great potential to be used as an effective photocatalyst, as already demonstrated in Ref.~\onlinecite{Chen2011}.   

Detailed high-resolution transmission electron microscopy (TEM) analysis\cite{Chen2011} of black TiO$_2$ revealed that the centers of the nanocrystals remain in the anatase TiO$_2$ structure, while only the surfaces of the crystals are modified.  However the surface is not completely disordered since after hydrogenation the Raman spectrum\cite{Chen2011} shows relatively sharp additional peaks in black TiO$_2$, even though the overall Raman spectrum is broader.  Additional Raman peaks were also found in subsequent studies\cite{Jiang2012,Zheng2012} and they could not be associated with any other known polymorph of TiO$_2$.

The presence of relatively sharp Raman peaks suggests that the hydrogenation might have induced a coherent structural distortion that is periodically repeating on the nanocrystal surface,rather than producing random disorder.  Even though extensive electronic and structural measurements have been done on black TiO$_2$, it remains unknown which structural change --- induced by hydrogenation under pressure --- might be responsible for the valence band energy shift. 

This is the question we address here from a theoretical point of view: {\it Can structural deformation alone move oxygen $p$-like valence states of TiO$_2$ to a higher energy and, if the answer is yes, how could such a deformation be stabilized?}

To answer these questions we introduce an approach that searches over a high-dimensional space of possible coherent structural distortions of TiO$_2$ anatase and selects  coherent structural distortions with a desired electronic band structure.  To our surprise, we found a single coherent structural distortion --- within a certain subspace --- that moves the valence band of TiO$_2$ to a higher energy without affecting the conduction band minimum.  Since the structural modification in black TiO$_2$ is likely not completely random --- as discussed before ---  our focus was on distortions with a small repetition period.

Near the end of the paper in Sec.~\ref{sec:stability} we also discuss two ways to stabilize this structure in a material, and how one of them might relate to the structure of black TiO$_2$.  We note here that numerous mechanisms have already been suggested to explain the color of black TiO$_2$.  In addition to the structural disorder mechanism, there is also a group of mechanisms that assigns the origin of the color to the modified chemistry of the nanocrystal.  Some suggestions relate to the presence of Ti$^{3+}$ ions, oxygen vacancies, or Ti--H groups. However, not all experiments done on black TiO$_2$ are using samples synthesized in the same way so it is not surprising that these different chemistries are not seen in all black TiO$_2$ samples.  The synthesis processes differ in three main aspects:  morphology of the TiO$_2$ nanostructures, chemical process used to synthesize TiO$_2$, as well as exposure to varying gaseous atmospheres.  Therefore, it is possible that same mechanism needs not explain black color in all samples. We refer the reader to Ref.~\onlinecite{Chen2015} and references within for an overview of proposed mechanisms and various synthesis routes.

While our theoretical approach was applied here only to the case of anatase-TiO$_2$ we believe that it is a quite general approach that can be applied to other materials as well, such as black WO$_3$.\cite{C2EE03158B}

There are numerous computational studies that deal with electronic structure modification of TiO$_2$ due to chemical dopants or oxygen vacancies.  For example, Ref.~\onlinecite{Aschauer2012} discusses adsorption of hydrogen into anatase surface and interplay with oxygen vacancy, while Ref.~\onlinecite{Raghunath} compares adsorption of atomic and molecular hydrogen.  Very few studies focus on the effect of coherent structural distortions alone.  One example is Ref.~\onlinecite{Liu2013} that focuses on the role of absorbing hydrogen atoms onto TiO$_2$ anatase nanocrystals and subsequent changes to the structure, while Ref.~\onlinecite{Yin2010} focuses on the role of external stress onto the electronic gap of bulk TiO$_2$. In both Ref.~\onlinecite{Liu2013} and  \onlinecite{Yin2010}, the authors find strong modification of the electron band gap when atoms are distorted along the tetragonal $c$-axis.

\subsection{Structure of bulk TiO$_2$ anatase}

We start by describing the crystal structure of TiO$_2$ anatase.  Its space group is body-centered tetragonal $I4_1/amd$.  There are two titanium and four oxygen atoms in the primitive cell and they occupy Wyckoff orbits $4b$ and $8e$.  Two Ti atoms are mapped to each other by a $4_1$ screw axis around the tetragonal $c$ axis.  

\begin{figure}[!h]
\centering
\includegraphics{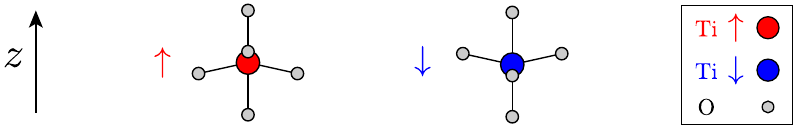}
\caption{\label{fig:str_2_atoms}Two kinds of Ti-O octahedra in bulk TiO$_2$ anatase.}
\end{figure}

The anatase structure consists of a network of edge-shared Ti-O octahedra that are slightly deformed. The octahedra consist of a Ti atom in its middle and six O atoms at its corners.  The deformation of Ti-O octahedra can be generated by starting from an ideal octahedron and then displacing four in-plane oxygen atoms along the $\pm z$ axis (tetragonal $c$ axis) so that two opposing oxygen atoms are displaced along $+z$ with the other two along $-z$. Depending on the choice of a pair of O atoms that are displaced along $+z$ or $-z$ one can construct two symmetry related octahedral environments.  These two choices of octahedral deformation are shown in Fig.~\ref{fig:str_2_atoms} and we will denote them symbolically as \kB and \kA.

\begin{figure}[!h]
\centering
\includegraphics{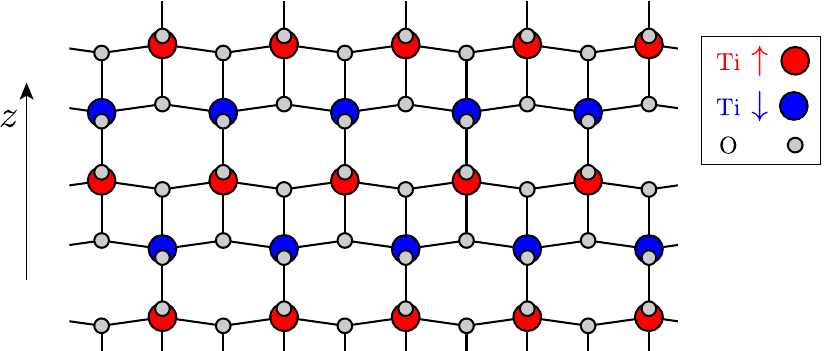}
\caption{\label{fig:initial} Crystal structure of pristine bulk TiO$_2$ anatase.}
\end{figure}

Now we will describe the three-dimensional arrangement of these octahedra in the anatase phase of TiO$_2$. As mentioned earlier, there are two Ti atoms in the primitive unit cell.  Octahedral environments around these Ti atoms are not the same, as they are related by a 4-fold screw axis which clearly maps \kB into \kA and vice versa.  Therefore, TiO$_2$ anatase crystal structure consists of planes of \kB and \kA octahedra alternating along the $z$ axis.  This structure is shown in Fig.~\ref{fig:initial} in a projection onto the $x$-$z$ plane.
Figure~\ref{fig:initial} clearly shows that  neighboring octahedral distortions are constrained to point in opposite directions along the $z$-axis and the same direction in the perpendicular $x$-$y$ plane.  We note that this constraint is imposed by the edge-sharing connectivity in the TiO$_2$ anatase and does not reduce space group symmetry.  In other words structure in which planar oxygens lie flat in the same plane as titanium atom would have the same space group symmetry. 

From the chemical bonding point of view these octahedral deformations originate from the fact that the oxygen atoms are surrounded by three titanium atoms which then induces some tendency for $sp^2$-like bonds on the oxygen atoms.

Symbolically we will denote the structure of anatase along the $z$-direction as,
\begin{center}
\ldots 
\kA \kB \kA \kB \kA \kB \kA \kB \kA \kB \kA \kB 
\ldots
\end{center}

\section{Approach}
\label{sec:bulk}

We now describe our approach based on density functional theory and GW calculations to find a simple structural modification of the TiO$_2$ anatase that will move the top of the valence band to higher energy while keeping the bottom of the conduction band nearly intact.  We divided this process  into two steps.  In the first step, described in this section, we find a bulk distortion of ideal TiO$_2$ anatase with the required electronic structure.  In the second step, described in Sec.~\ref{sec:stability}, we  discuss possible ways to stabilize this distortion in a real material, such as black TiO$_2$.

For structural calculations we use density functional theory calculation based on Perdew-Burke-Ernzerhof\cite{pbe1997} (PBE) approximation to the exchange-correlation functional.  We use GPAW\cite{PhysRevB.71.035109} and Quantum-ESPRESSO\cite{giannozzi} computer packages. Both packages give very similar results for TiO$_2$ anatase (and many other materials~\cite{Lejaeghere2016}).  For the Quantum-ESPRESSO calculations, we used a plane-wave basis with a cutoff energy of 60~Ry and 600~Ry for the electron wavefunction and density, respectively.  For calculations using the GPAW code, we used a real-space grid with $h=0.18$~\AA\ grid spacing.  In both cases, we use a 4x4x4 k-grid and we include semi-core states on the titanium atoms. To obtain an accurate electronic band structure, we used the GW approximation as implemented in the BerkeleyGW package.\cite{PhysRevB.34.5390,Deslippe20121269} The GW calculation is well converged with 1,000 empty states per single TiO$_2$ formula unit (in all cases) and with a dielectric matrix cutoff energy of 40~Ry.

Within the density functional theory framework one conventionally determines the ground-state crystal structure by minimizing the total energy $E_{\rm tot}$ over some structural parameters $\xi_i$.  Here by $\xi_i$ we will denote both internal atom displacements as well as changes to the lattice vectors.  

In this work we will instead start from the fully relaxed TiO$_2$ anatase structure and then minimize the following function,
\begin{align}
  F(\xi_i) = E_{\rm tot}(\xi_i) + \lambda \left[ 
  	- \epsilon_{\rm v}(\xi_i) + \left| \epsilon_{\rm c}(\xi_i) - \epsilon_{\rm c}(0) \right| \right].
	\label{eq:F}	
\end{align}
Here vector $\xi_i$ describes distortions away from the ideal anatase phase and the Kohn-Sham eigenvalues ($\epsilon_{\rm c}$ and $\epsilon_{\rm v}$) are in principle measured with respect to the vacuum level.  With $\lambda > 0$, the second term in Eq.~\eqref{eq:F} rewards distortions that move the valence band maximum $\epsilon_{\rm v}$ to higher energies and penalizes those that move the conduction band minimum $\epsilon_{\rm c}$ relative to that of the ideal anatase, $\epsilon_{\rm c}(0)$.  The first term in Eq.~\eqref{eq:F} is the total energy (per one atom), as in the conventional structural relaxation.  The role of the first term is to ensure that the band-engineering enforced by the second term does not change the total energy by too much. 

Vector $\xi_i$ includes all 21 distortions that do not change the number of atoms in the primitive unit cell of TiO$_2$ anatase.  These include $6 \times 3 - 3 = 15$ internal degrees of freedom (there are $6$ atoms in the primitive cell and we do not count rigid translations) and they also include $3 \times 3 - 3 = 6$ lattice vector distortions (we do not count rigid rotations).  Typically these distortions can be decomposed into irreducible representations of the space group of TiO$_2$ anatase.

For computational simplicity in Eq.~\eqref{eq:F} we measure band energies $\epsilon_{\rm c}$, $\epsilon_{\rm v}$ relative to the potential in which the average electrostatic (Hartree) potential in the unit cell equals zero.\footnote{The definition of the average electrostatic potential depends on the volume of the unit cell. Therefore, strictly speaking, one should not allow lattice vectors to change during minimization of $F$. However, in our calculations we find that resulting structure that minimizes $F$ does not depend strongly on whether lattice vectors are allowed to change or not.}  Another possible simplification would be to measure those states relative to the core or semi-core states.  However, we find that in our calculation with zeroed out average Hartree potential, the Ti $3s$ and $3p$ semi-core states change by at most 20~meV, so both approximate approaches give nearly the same $\xi_i$.  We later confirmed the validity of this approach by performing an explicit surface calculation and computing the band structure relative to the vacuum level above the surface.

The dimensionless scalar $\lambda$ appearing in Eq.~\eqref{eq:F} measures the importance of the second term relative to the first term. With $\lambda=0$ minimizing $F$ is the same as minimizing $E_{\rm tot}$ while with $\lambda=+\infty$ band-engineering is done without regard for the increase in the total energy $E_{\rm tot}$.  We used $\lambda=1.7$ and confirmed that  $\lambda=0.3$ and $\lambda=0.8$ give nearly the same optimized vector $\xi$ up to a constant prefactor.

We minimized $F$ with a Nelder-Mead simplex algorithm\cite{Nelder} since that algorithm does not need derivatives of $F$ with respect to $\xi_i$.  While computing derivative of $E_{\rm tot}$ with respect to $\xi_i$ would have been relatively straightforward, obtaining derivatives of the eigenvalues $\epsilon_{\rm c}$ and $\epsilon_{\rm v}$ is computationally more intensive.  Therefore we rely here on minimization algorithm that doesn't need derivatives of $F$.  

We also note here that there is no efficient, general numerical algorithm to find a global minimum of a high-dimensional function (such as $F$).  A local optimization algorithm such as Nelder-Mead simplex algorithm used in this paper may find a local minimum --- but not necessarily a global minimum --- of the function $F$.  We initialized the minimization algorithm at the fully relaxed ground state structure of TiO$_2$ anatase.  Initial search simplex is composed of individual distortions of all 21 degrees of freedom, and the length of each side of the simplex is initially set at 0.025~\AA.  In the subsequent steps this simplex is  adaptively increased or decreased in length following Ref.~\onlinecite{Nelder} with standard values of the so-called reflection, contraction, and expansion coefficients ($\alpha=1$, $\beta=1/2$, and $\gamma=2$, respectively).

\begin{figure}[!h]
\centering
\includegraphics{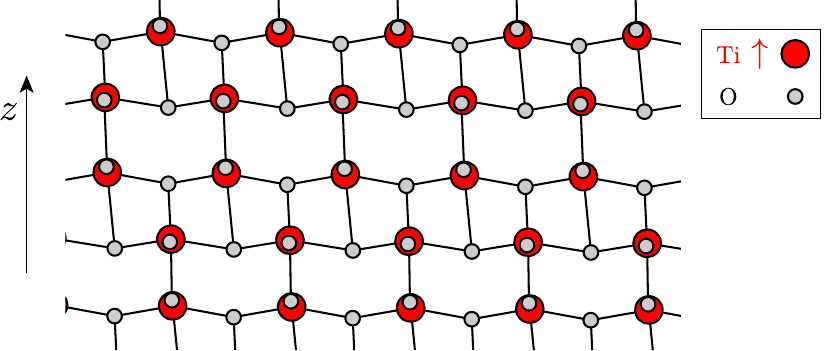}
\caption{\label{fig:final} Structural distortion $\xi_i$ that reduces value of $F(\xi_i)$.}
\end{figure}

The resulting calculated structure that minimizes $F$ can schematically be represented as,
\begin{center}
\ldots 
\kB \kB \kB \kB \kB \kB \kB \kB \kB \kB \kB \kB
\ldots
\end{center}
or 
\begin{center}
\ldots 
\kA \kA \kA \kA \kA \kA \kA \kA \kA \kA \kA \kA
\ldots
\end{center}
In other words, TiO$_2$ anatase will have optimal band alignment for water splitting when all of its oxygen octahedral distortions point in the same direction.  Projection of the structure on the $x$-$z$ plane is shown in Fig.~\ref{fig:final}.
As can be seen from the figure, half of the Ti-O bonds along the $z$-axis are stretched by this distortion, which is likely responsible for the majority of the energy penalty as well as the shift in the valence band maximum.  As mentioned earlier, sensitivity of the band structure on displacements along the $c$ axis was discussed before in Refs.~\onlinecite{Liu2013,Yin2010}. 

We also note here that distortion for half of the octahedrons is not perfectly represented by symbol \kB since atoms O--Ti--O are nearly collinear along the $y$ direction for those octahedrons.  However, distortion of those atoms still point along the same direction, and the amount of distortion in the structure is linearly proportional with chosen value of $\lambda$.

\begin{figure}
\centering
\includegraphics{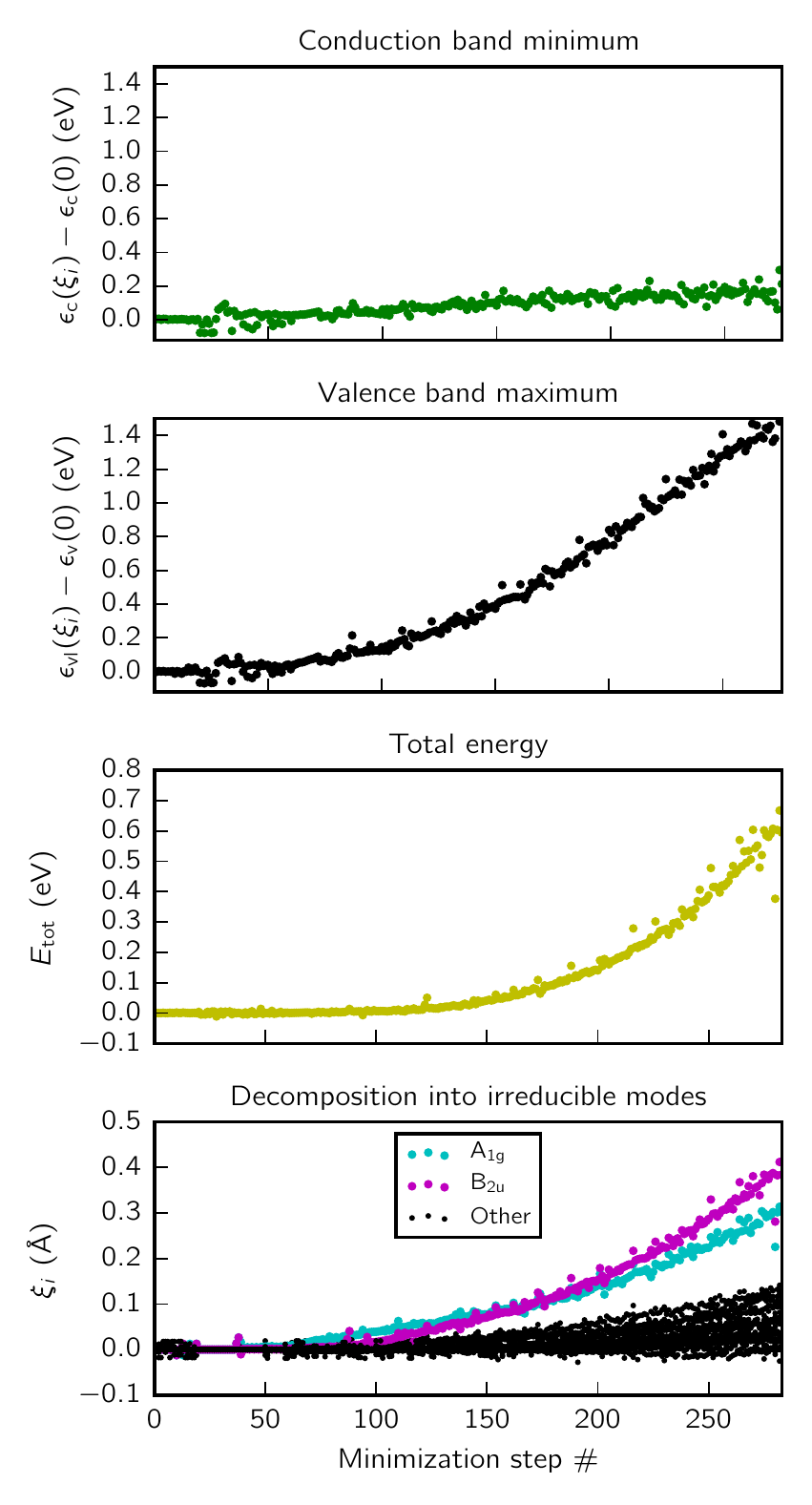}
\caption{\label{fig:brute} Conduction band minimum ($\epsilon_{\rm c}$), valence band maximum ($\epsilon_{\rm v}$), the total energy ($E_{\rm tot}$), and decomposition into ireducible modes ($\xi_i$) along a path that minimizes $F$.}
\end{figure}

The top three panels of Fig.~\ref{fig:brute} show the values of $\epsilon_{\rm c}$, $\epsilon_{\rm v}$, and $E_{\rm tot}$ along the calculated path that minimizes $F$.  All three quantities are shown relative to their values in an undistorted structure ($\xi_i=0$).  As can be seen from Fig.~\ref{fig:brute} conduction band minimum is kept nearly constant while valence band maximum moves up by about 1.4~eV.  The total energy of the system is increased by 0.6~eV per atom at the optimal structure.  As mentioned earlier, $E_{\rm tot}$, $\epsilon_{\rm c}$, $\epsilon_{\rm v}$, and $\xi_i$ scale nearly linearly with chosen value of $\lambda$.

\begin{figure*}
\centering
\includegraphics{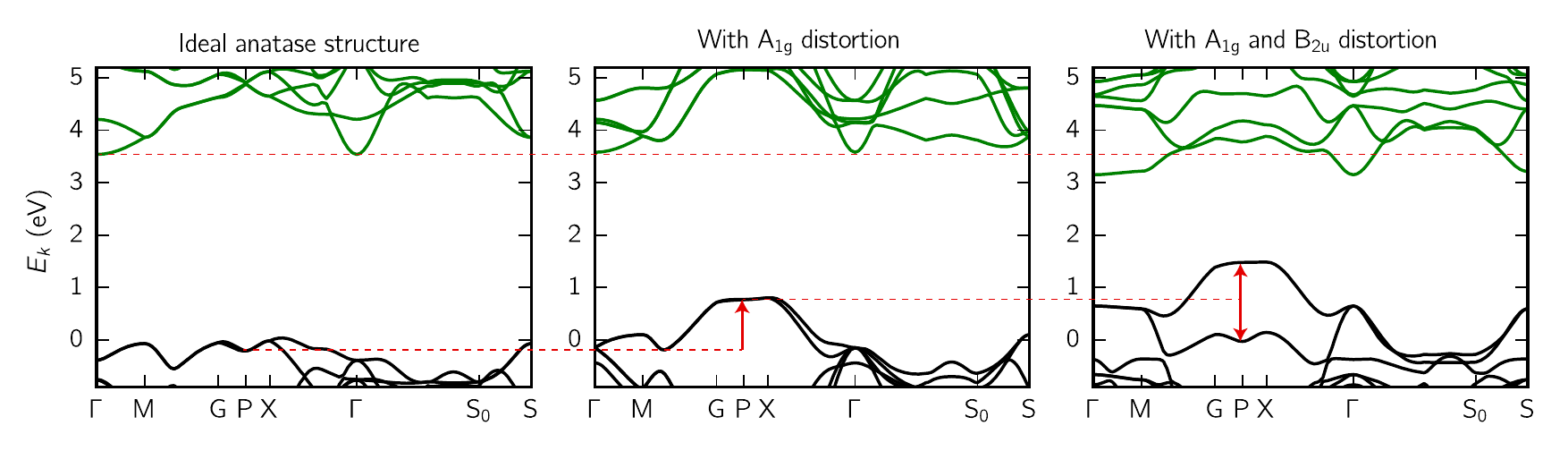}
\caption{\label{fig:bulk_bs}Band structure along a path in $k$-space in pristine TiO$_2$ anatase (left), with applied $A_{1\rm g}$ distortion (middle), and with both $A_{1\rm g}$ and $B_{2\rm u}$ distortions (right). The titanium core levels are aligned among three panels.  The red dashed lines are guides to the eye.}
\end{figure*}

The bottom panel of Fig.~\ref{fig:brute} shows a decomposition of distortions along the minimization path in terms of irreducible representations.  It is clear that there are two irreducible distortions that dominate; both involve displacements of the oxygen atoms whose symmetry transforms according to the irreducible representations of $A_{1 \rm g}$ and $B_{2\rm u}$.  While the first mode preserves space group symmetry the $B_{2\rm u}$ mode breaks the inversion symmetry.  We now discuss the effect of these two modes on the band structure of TiO$_2$ anatase. The effect of $A_{1 \rm g}$ and $B_{2 \rm u}$ is shown in Fig.~\ref{fig:bulk_bs}.
By comparing the left and middle panels of the figure one can see that the $A_{1 \rm g}$ mode moves up in energy oxygen valence bands near the $X$ point.  It has nearly no effect on the titanium conduction bands.   If we include the effect of the symmetry breaking mode $B_{2\rm u}$ (right panel) one can see that this mode splits the double degenerate valence states near the $X$ point with a  minor effect on the conduction bands.  The main effect of distortions belonging to other irreducible representations was found to compensate for this minor effect on the conduction bands.

The band structures shown in Figure~\ref{fig:bulk_bs} are scissor shifted by 1.39~eV.  We obtained this value by computing the GW band structure of pristine TiO$_2$ anatase and confirming that the main effect of GW self-energy in this material is to scissor shift bands by that amount (1.39~eV).

\section{Structural stabilization}
\label{sec:stability}

\begin{figure}[!b]
\centering
\includegraphics{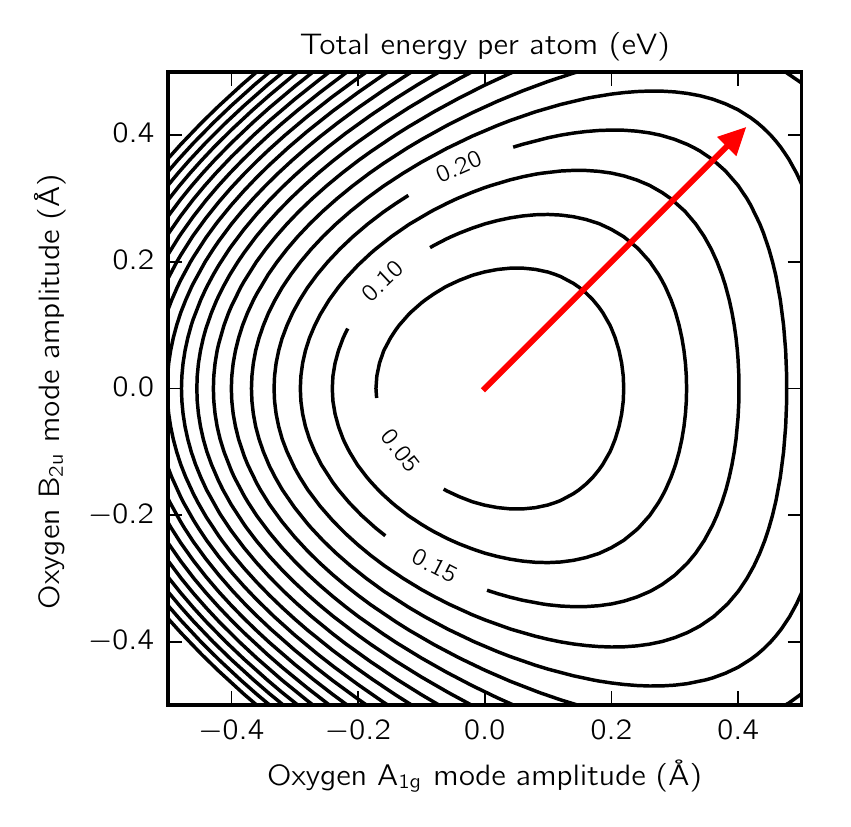}
\caption{\label{fig:landscape} Contour plot of total energy $E_{\rm tot}$ per one atom as a function of distortions $A_{1\rm g}$ and $B_{2\rm u}$.  Red arrow shows approximate direction in which $F$ is minimized.}
\end{figure}

The distortion \kB\kB\kB\kB discussed in earlier section has the desired electronic properties but it is structurally unstable since it minimizes $F$ instead of the true total energy $E_{\rm tot}$.  We confirmed that this distortion is structurally unstable by starting from the structure obtained with $\lambda=1.7$ and then proceeding with the conventional structural minimization.  The structure quickly converges to the pristine anatase TiO$_2$.  To further confirm that this structure is unstable we explored the energy landscape spanned by two dominant distortions $A_{1 \rm g}$ and $B_{2 \rm u}$.  As shown in Fig.~\ref{fig:landscape} the total energy has a single global minimum when both distortions are zero without any sign of alternative local minima.  The red arrow in this figure indicates approximate direction along which $F$ is minimized.  We also confirmed that imposing uniaxial, biaxial, or isotropic strain cannot stabilize this structure. We suspect that the \kB\kA\kB\kA arrangement is so robust since it is dictated by the connectivity of oxygen octahedra and not by symmetry breaking.  Therefore, as long as connectivity remains the same, \kB\kA\kB\kA will remain to be the ground-state.

Driven by this insight in the following two subsections we propose two ways to stabilize the parallel orientation of octahedral distortions, one in bulk and one at the surface.

\subsection{Stabilization in the bulk}

To construct parallel orientation of the neighboring Ti-O octahedra let us start from a pristine configuration of TiO$_2$ anatase,
\begin{center}
\ldots 
\kA \kB \kA \kB \kA \kB \kA \kB \kA \kB \kA \kB
\ldots
\end{center}
As step 1 we cut this infinite sample in two halves,
\begin{center}
\ldots 
\kA \kB \kA \kB \kA \kB \quad\quad \kA \kB \kA \kB \kA \kB
\ldots
\end{center}
in step 2 we rotate the right half of the sample by $90^{\circ}$ around the $z$-axis.  This rotation flips \kB to \kA and vice versa,
\begin{center}
\ldots 
\kA \kB \kA \kB \kA \kB \quad\quad \kB \kA \kB \kA \kB \kA 
\ldots
\end{center}
Finally, in step 3 we bring back together two halves of the sample 
\begin{center}
\ldots 
\kA \kB \kA \kB \kA \kB \kB \kA \kB \kA \kB \kA
\ldots
\end{center}
which leaves a desired parallel configuration of neighboring octahedra in the middle of the sample.  This kind of planar defect is commonly referred to as a twin boundary.  We note that one could have arrived at the same structure by individually translating right half of the Ti--O planes in the direction perpendicular to the $z$-axis and then displacing oxygen atoms along the $z$-axis so that \kB flips to \kA and vice versa.

In the pristine anatase there are two Ti--O bonds per unit cell for each \kA\kB pair of octahedra.  However, \kB\kB pair allows for only a single Ti--O bond per cell.  Therefore while in step 2 we broke two bonds only one of them was recovered in step 3.  Nevertheless, we expect that this configuration is stable since removing \kB\kB from the middle would require half of the sample to flip from \kB to \kA and vice versa.  This process would naturally involve a large activation barrier.  Therefore if \kB\kB configuration is formed in a crystal during its growth, we expect it to remain in the crystal.

\begin{figure}[!h] 
\centering
\includegraphics{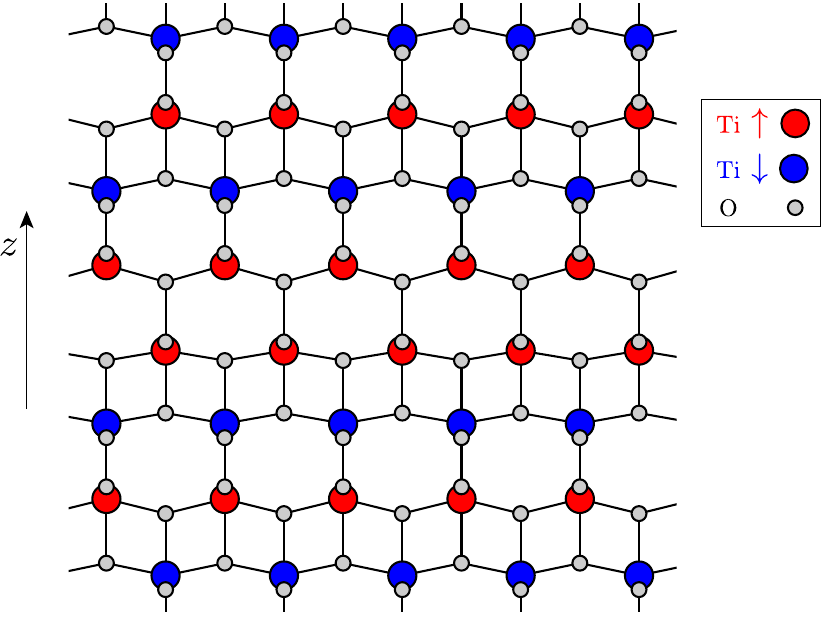}
\caption{\label{fig:dw_struc} Stable structure with \kB\kB configuration in bulk.  This distortion is also stable on the surface.}
\end{figure}

Indeed, we confirmed that this kind of structure is stable even after full structural relaxation within the density functional theory.  The relaxed structure is shown in Fig.~\ref{fig:dw_struc}. 
We also checked dynamical stability of this configuration by computing the phonon frequencies at the zone center and confirming that there are no $\Gamma$ soft phonon modes in the structure. 
 
We also calculated the band structure of such a configuration within the GW approach.  The calculated indirect band gap is reduced from 3.5~eV in pristine TiO$_2$  to 2.0~eV in TiO$_2$ with \kB\kB configuration.  This calculation was done in a super-cell that contains 15 layers of TiO$_2$ planes.  We obtained nearly the same gap (2.1~eV) if we perform the slab calculation within DFT and then use the scissor shift computed for the pristine TiO$_2$ anatase.  This calculation confirms that the scissor shift is a good approximation for the GW calculation of TiO$_2$ anatase.

\subsection{Stabilization on the surface}

While parallel configuration of octahedral deformations is stable in bulk, it is less clear whether this configuration is stable on the surface of a crystal.  However, detailed study of various surfaces would require comparison of surface energies of many surface terminations with the possible reconstructions in the presence of hydrogen atoms.  Therefore, we focus here for simplicity only on the most natural surface termination for our study.  The most natural surface is $(001)$ since the \kB\kB structural motif lies in that plane.  We leave discussion of other surfaces for future studies.

\begin{figure}[!h] 
\centering
\includegraphics{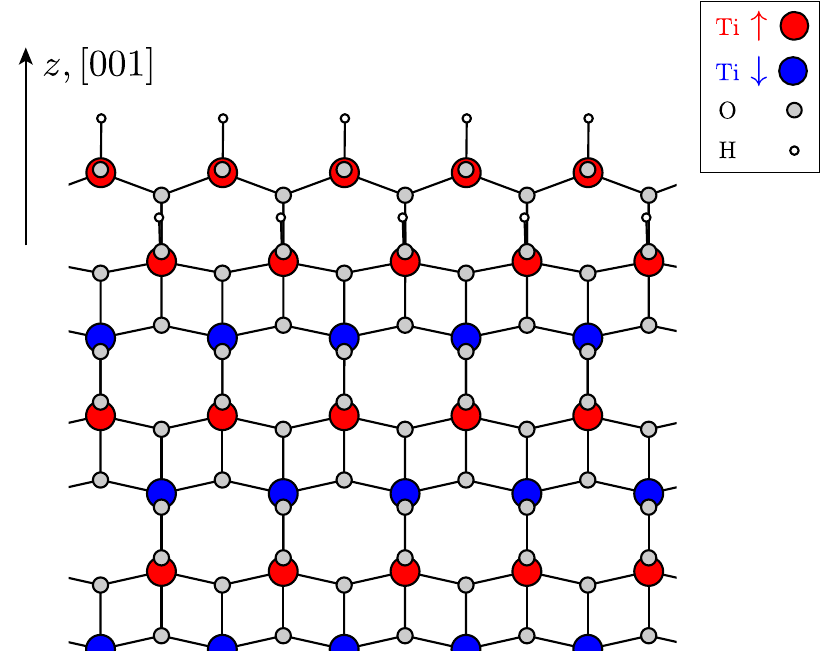}
\caption{\label{fig:surface_struc} Surface with \kB\kB termination stabilized using a  simplified approach based on pseudohydrogens (see text for details).}
\end{figure}

Surfaces of partially covalent materials such as TiO$_2$ typically contain dangling bonds that can be passivated by hydrogen atoms.  As mentioned earlier, pristine TiO$_2$ anatase contains two Ti-O bonds along the $z$-axis per surface unit cell.  Therefore $(001)$ surface of pristine TiO$_2$ anatase will create two dangling bonds per surface unit cell.  However, with parallel configuration of octahedral deformations at the surface,
\begin{center}
\ldots 
\kA \kB \kA \kB \kA \kB \kA \kB \kB \ [vacuum]
\end{center}
number of dangling bonds increases from two to four as \mbox{\kB \kB} configuration breaks one of the Ti--O bonds (see Fig.~\ref{fig:surface_struc}). Therefore we expect that at high enough hydrogen pressure and temperature the free energy of the \mbox{\kB \kB} configuration at the surface will be lower than that of a regular TiO$_2$ anatase surface.  This is consistent with the fact that the absorption energy of hydrogen is quite high (about 2~eV).\cite{Aschauer2012}

We note that this configuration on the surface can be achieved by a translation of the top-most TiO$_2$ by $(\frac{1}{2} \frac{1}{2} 0)$ and then slight displacement of oxygen atoms along the $z$ axis to shift \kA into \kB.

\begin{figure}[!t]
\centering
\includegraphics{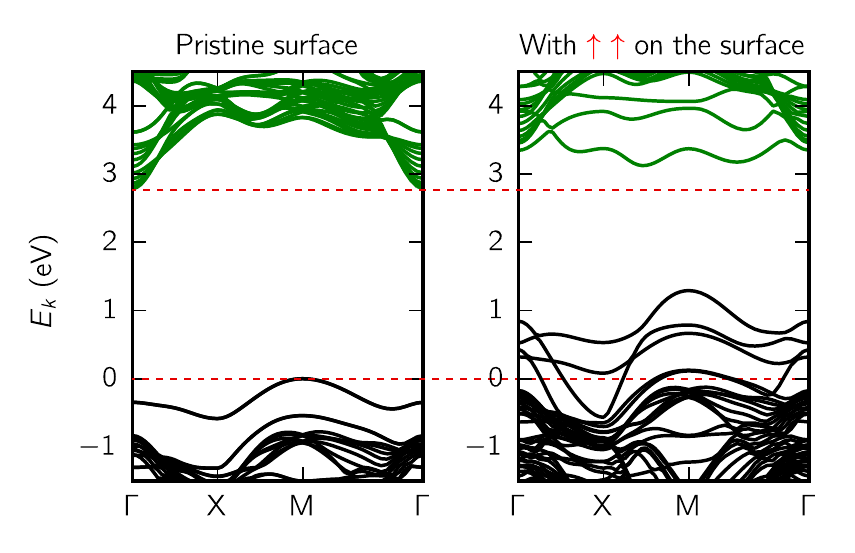}
\caption{\label{fig:surface_bs} Band structure without (left) and with (right) \kB\kB on the surface.  The band structure is aligned relative to the vacuum level above each surface. The red dashed lines are guides to the eye.}
\end{figure}

We performed a full structural relaxation of the TiO$_2$ surface terminated by \kB \kB and find it to be in a structural local minimum of energy.  The band structure of the fully relaxed structure is shown in Fig.~\ref{fig:surface_bs} along with the band structure of the pristine anatase-TiO$_2$ surface.  We aligned the band structures of these two calculations so that the explicitly calculated vacuum levels match.  As can be clearly seen from the figure, surface terminated by \kB \kB has a valence band maximum that is higher by about 1.3~eV.

The full surface calculation in the presence of hydrogen is computationally expensive both due to the presence of surface reconstruction and due to the possible role of oxygen vacancies,\cite{Aschauer2012} and it will thus be left for future study.  However, we performed a simplified surface calculation by passivating surface dangling bonds by pseudohydrogen, following Ref.~\onlinecite{PhysRevB.72.125325}.  We find that pristine anatase-TiO$_2$ $(001)$ surface can be passivated by replacing each missing Ti atom with a pseudohydrogen with $4/3$ charge and each missing O atom with $2/3$ charged pseudohydrogen.  In the case of \kB \kB terminated surface, we can passivate the surface with the same pseudhydrogen atoms.  However, in this case we find that the pseudohydrogen passivating oxygen atom prefers to sit on the second top-most surface layer of atoms, as shown in Fig.~\ref{fig:surface_struc}.  This configuration then further stabilizes the structure as it prevents the top-most layer of TiO$_2$ from sliding back and thus eliminating \kB \kB termination.

\subsection{Candidate structure for black TiO$_2$}

We now discuss our candidate structure for black TiO$_2$ nano-sized powder.\cite{Chen2011}  As is understood from the high-resolution TEM data, the core of the TiO$_2$ nano-particles remains in a highly crystalline anatase TiO$_2$ state, while the surface of the crystal has a different structure that is disordered but not amorphous.  Furthermore, the structural modification at the surface of the nanocrystal was induced by exposure to hydrogen gas under high pressure (20~bar) and temperature (200$^{\circ}$C).

Therefore, it is plausible that the high pressure and temperature modified the pristine surface to the one with parallel arrangement of oxygen octahedra, as discussed in the earlier section.  We note that this kind of deformation requires only a slight in-plane translation of the TiO$_2$ layer on the surface. 

Of course, it is possible that some other structure might be responsible for blackness of TiO$_2$ powder reported in Ref.~\onlinecite{Chen2011}.  Nevertheless, the structural motif that we found (\kB\kB) is unique in the space of distortions that do not enlarge the number of atoms in the anatase-TiO$_2$ unit cell.  Therefore, within this space this is the simplest structural distortion of TiO$_2$ anatase that achieves the desired electronic structure modification.

We note here that there are physical consequences of the \kB\kB structural motif which agree with experiment in addition to raising the energy of the valence band maximum.   For example, this structural motif breaks the inversion symmetry and thus activates some Raman modes that are inactive in the pristine TiO$_2$ anatase.  In addition, this new distorted motif is not amorphous but instead contains {\it ordered} structure with translational symmetry and hence well defined Raman peaks.  This is consistent with the fact that there are several Raman modes in black-TiO$_2$ that are not present in the pristine anatase TiO$_2$.\cite{Chen2011} Finally, this structural motif splits the doubly degenerate bands near the $X$ point which means that its absorption edge will have a two-legged feature, which is again consistent with the experiment.\cite{Chen2011}

\section{Outlook}

We presented a general strategy to band-engineer a wide range of materials by exploring the large subspace of coherent structural distortions of a bulk crystal structure.  Furthermore, we applied this approach to the anatase-TiO$_2$ and found a candidate structure for the so-called black-TiO$_2$.  We hope that this computational approach might be used to find alternative structures in other materials with desirable band structures for applications.

\begin{acknowledgments}
This work was supported the Theory of Materials Program at the Lawrence Berkeley National Lab, funded by the Director, Office of Science, Office of Basic Energy Sciences, Materials Sciences and Engineering Division, U.S. Department of Energy under Contract No. DE-AC02-05CH11231. Computational resources have been provided by the DOE at Lawrence Berkeley National Laboratory's NERSC facility. S.S. acknowledges support from the MEXT Japan Elements Strategy Initiative to Form Core Research Center, and JSPS KAKENHI Grant No.JP25107005. Y.A. acknowledge support from JSPS Grant No.JP14J11856.
\end{acknowledgments}

\bibliography{pap}

\begin{thebibliography}{25}%
\makeatletter
\providecommand \@ifxundefined [1]{%
 \@ifx{#1\undefined}
}%
\providecommand \@ifnum [1]{%
 \ifnum #1\expandafter \@firstoftwo
 \else \expandafter \@secondoftwo
 \fi
}%
\providecommand \@ifx [1]{%
 \ifx #1\expandafter \@firstoftwo
 \else \expandafter \@secondoftwo
 \fi
}%
\providecommand \natexlab [1]{#1}%
\providecommand \enquote  [1]{``#1''}%
\providecommand \bibnamefont  [1]{#1}%
\providecommand \bibfnamefont [1]{#1}%
\providecommand \citenamefont [1]{#1}%
\providecommand \href@noop [0]{\@secondoftwo}%
\providecommand \href [0]{\begingroup \@sanitize@url \@href}%
\providecommand \@href[1]{\@@startlink{#1}\@@href}%
\providecommand \@@href[1]{\endgroup#1\@@endlink}%
\providecommand \@sanitize@url [0]{\catcode `\\12\catcode `\$12\catcode
  `\&12\catcode `\#12\catcode `\^12\catcode `\_12\catcode `\%12\relax}%
\providecommand \@@startlink[1]{}%
\providecommand \@@endlink[0]{}%
\providecommand \url  [0]{\begingroup\@sanitize@url \@url }%
\providecommand \@url [1]{\endgroup\@href {#1}{\urlprefix }}%
\providecommand \urlprefix  [0]{URL }%
\providecommand \Eprint [0]{\href }%
\providecommand \doibase [0]{http://dx.doi.org/}%
\providecommand \selectlanguage [0]{\@gobble}%
\providecommand \bibinfo  [0]{\@secondoftwo}%
\providecommand \bibfield  [0]{\@secondoftwo}%
\providecommand \translation [1]{[#1]}%
\providecommand \BibitemOpen [0]{}%
\providecommand \bibitemStop [0]{}%
\providecommand \bibitemNoStop [0]{.\EOS\space}%
\providecommand \EOS [0]{\spacefactor3000\relax}%
\providecommand \BibitemShut  [1]{\csname bibitem#1\endcsname}%
\let\auto@bib@innerbib\@empty
\bibitem [{\citenamefont {Fujishima}\ and\ \citenamefont
  {Honda}(1972)}]{Fujishima1972}%
  \BibitemOpen
  \bibfield  {author} {\bibinfo {author} {\bibfnamefont {A.}~\bibnamefont
  {Fujishima}}\ and\ \bibinfo {author} {\bibfnamefont {K.}~\bibnamefont
  {Honda}},\ }\href {\doibase 10.1038/238037a0} {\bibfield  {journal} {\bibinfo
   {journal} {Nature}\ }\textbf {\bibinfo {volume} {238}},\ \bibinfo {pages}
  {37} (\bibinfo {year} {1972})}\BibitemShut {NoStop}%
\bibitem [{\citenamefont {Pascual}\ \emph {et~al.}(1978)\citenamefont
  {Pascual}, \citenamefont {Camassel},\ and\ \citenamefont
  {Mathieu}}]{PhysRevB.18.5606}%
  \BibitemOpen
  \bibfield  {author} {\bibinfo {author} {\bibfnamefont {J.}~\bibnamefont
  {Pascual}}, \bibinfo {author} {\bibfnamefont {J.}~\bibnamefont {Camassel}}, \
  and\ \bibinfo {author} {\bibfnamefont {H.}~\bibnamefont {Mathieu}},\ }\href
  {\doibase 10.1103/PhysRevB.18.5606} {\bibfield  {journal} {\bibinfo
  {journal} {Phys. Rev. B}\ }\textbf {\bibinfo {volume} {18}},\ \bibinfo
  {pages} {5606} (\bibinfo {year} {1978})}\BibitemShut {NoStop}%
\bibitem [{\citenamefont {Xu}\ and\ \citenamefont {Schoonen}(2000)}]{Xu543}%
  \BibitemOpen
  \bibfield  {author} {\bibinfo {author} {\bibfnamefont {Y.}~\bibnamefont
  {Xu}}\ and\ \bibinfo {author} {\bibfnamefont {M.~A.}\ \bibnamefont
  {Schoonen}},\ }\href {\doibase 10.2138/am-2000-0416} {\bibfield  {journal}
  {\bibinfo  {journal} {American Mineralogist}\ }\textbf {\bibinfo {volume}
  {85}},\ \bibinfo {pages} {543} (\bibinfo {year} {2000})}\BibitemShut
  {NoStop}%
\bibitem [{\citenamefont {Asahi}\ \emph {et~al.}(2001)\citenamefont {Asahi},
  \citenamefont {Morikawa}, \citenamefont {Ohwaki}, \citenamefont {Aoki},\ and\
  \citenamefont {Taga}}]{Asahi269}%
  \BibitemOpen
  \bibfield  {author} {\bibinfo {author} {\bibfnamefont {R.}~\bibnamefont
  {Asahi}}, \bibinfo {author} {\bibfnamefont {T.}~\bibnamefont {Morikawa}},
  \bibinfo {author} {\bibfnamefont {T.}~\bibnamefont {Ohwaki}}, \bibinfo
  {author} {\bibfnamefont {K.}~\bibnamefont {Aoki}}, \ and\ \bibinfo {author}
  {\bibfnamefont {Y.}~\bibnamefont {Taga}},\ }\href {\doibase
  10.1126/science.1061051} {\bibfield  {journal} {\bibinfo  {journal}
  {Science}\ }\textbf {\bibinfo {volume} {293}},\ \bibinfo {pages} {269}
  (\bibinfo {year} {2001})}\BibitemShut {NoStop}%
\bibitem [{\citenamefont {Khan}\ \emph {et~al.}(2002)\citenamefont {Khan},
  \citenamefont {Al-Shahry},\ and\ \citenamefont {Ingler}}]{Khan2243}%
  \BibitemOpen
  \bibfield  {author} {\bibinfo {author} {\bibfnamefont {S.~U.~M.}\
  \bibnamefont {Khan}}, \bibinfo {author} {\bibfnamefont {M.}~\bibnamefont
  {Al-Shahry}}, \ and\ \bibinfo {author} {\bibfnamefont {W.~B.}\ \bibnamefont
  {Ingler}},\ }\href {\doibase 10.1126/science.1075035} {\bibfield  {journal}
  {\bibinfo  {journal} {Science}\ }\textbf {\bibinfo {volume} {297}},\ \bibinfo
  {pages} {2243} (\bibinfo {year} {2002})}\BibitemShut {NoStop}%
\bibitem [{\citenamefont {Devi}\ and\ \citenamefont
  {Kavitha}(2013)}]{Devi2013559}%
  \BibitemOpen
  \bibfield  {author} {\bibinfo {author} {\bibfnamefont {L.~G.}\ \bibnamefont
  {Devi}}\ and\ \bibinfo {author} {\bibfnamefont {R.}~\bibnamefont {Kavitha}},\
  }\href {\doibase http://dx.doi.org/10.1016/j.apcatb.2013.04.035} {\bibfield
  {journal} {\bibinfo  {journal} {Applied Catalysis B: Environmental}\ }\textbf
  {\bibinfo {volume} {140–141}},\ \bibinfo {pages} {559 } (\bibinfo {year}
  {2013})}\BibitemShut {NoStop}%
\bibitem [{\citenamefont {Choi}\ \emph {et~al.}(1994)\citenamefont {Choi},
  \citenamefont {Termin},\ and\ \citenamefont
  {Hoffmann}}]{doi:10.1021/j100102a038}%
  \BibitemOpen
  \bibfield  {author} {\bibinfo {author} {\bibfnamefont {W.}~\bibnamefont
  {Choi}}, \bibinfo {author} {\bibfnamefont {A.}~\bibnamefont {Termin}}, \ and\
  \bibinfo {author} {\bibfnamefont {M.~R.}\ \bibnamefont {Hoffmann}},\ }\href
  {\doibase 10.1021/j100102a038} {\bibfield  {journal} {\bibinfo  {journal}
  {The Journal of Physical Chemistry}\ }\textbf {\bibinfo {volume} {98}},\
  \bibinfo {pages} {13669} (\bibinfo {year} {1994})}\BibitemShut {NoStop}%
\bibitem [{\citenamefont {Chen}\ \emph {et~al.}(2011)\citenamefont {Chen},
  \citenamefont {Liu}, \citenamefont {Yu},\ and\ \citenamefont
  {Mao}}]{Chen2011}%
  \BibitemOpen
  \bibfield  {author} {\bibinfo {author} {\bibfnamefont {X.}~\bibnamefont
  {Chen}}, \bibinfo {author} {\bibfnamefont {L.}~\bibnamefont {Liu}}, \bibinfo
  {author} {\bibfnamefont {P.~Y.}\ \bibnamefont {Yu}}, \ and\ \bibinfo {author}
  {\bibfnamefont {S.~S.}\ \bibnamefont {Mao}},\ }\href {\doibase
  10.1126/science.1200448} {\bibfield  {journal} {\bibinfo  {journal}
  {Science}\ }\textbf {\bibinfo {volume} {331}},\ \bibinfo {pages} {746}
  (\bibinfo {year} {2011})}\BibitemShut {NoStop}%
\bibitem [{\citenamefont {Jiang}\ \emph {et~al.}(2012)\citenamefont {Jiang},
  \citenamefont {Zhang}, \citenamefont {Jiang}, \citenamefont {Rong},
  \citenamefont {Wang}, \citenamefont {Wu},\ and\ \citenamefont
  {Pan}}]{Jiang2012}%
  \BibitemOpen
  \bibfield  {author} {\bibinfo {author} {\bibfnamefont {X.}~\bibnamefont
  {Jiang}}, \bibinfo {author} {\bibfnamefont {Y.}~\bibnamefont {Zhang}},
  \bibinfo {author} {\bibfnamefont {J.}~\bibnamefont {Jiang}}, \bibinfo
  {author} {\bibfnamefont {Y.}~\bibnamefont {Rong}}, \bibinfo {author}
  {\bibfnamefont {Y.}~\bibnamefont {Wang}}, \bibinfo {author} {\bibfnamefont
  {Y.}~\bibnamefont {Wu}}, \ and\ \bibinfo {author} {\bibfnamefont {C.-x.}\
  \bibnamefont {Pan}},\ }\href {http://pubs.acs.org/doi/abs/10.1021/jp307573c}
  {\bibfield  {journal} {\bibinfo  {journal} {J.Phys.Chem.C}\ }\textbf
  {\bibinfo {volume} {116}},\ \bibinfo {pages} {22619} (\bibinfo {year}
  {2012})}\BibitemShut {NoStop}%
\bibitem [{\citenamefont {Zheng}\ \emph {et~al.}(2012)\citenamefont {Zheng},
  \citenamefont {Huang}, \citenamefont {Lu}, \citenamefont {Wang},
  \citenamefont {Qin}, \citenamefont {Zhang}, \citenamefont {Dai},\ and\
  \citenamefont {Whangbo}}]{Zheng2012}%
  \BibitemOpen
  \bibfield  {author} {\bibinfo {author} {\bibfnamefont {Z.}~\bibnamefont
  {Zheng}}, \bibinfo {author} {\bibfnamefont {B.}~\bibnamefont {Huang}},
  \bibinfo {author} {\bibfnamefont {J.}~\bibnamefont {Lu}}, \bibinfo {author}
  {\bibfnamefont {Z.}~\bibnamefont {Wang}}, \bibinfo {author} {\bibfnamefont
  {X.}~\bibnamefont {Qin}}, \bibinfo {author} {\bibfnamefont {X.}~\bibnamefont
  {Zhang}}, \bibinfo {author} {\bibfnamefont {Y.}~\bibnamefont {Dai}}, \ and\
  \bibinfo {author} {\bibfnamefont {M.-H.}\ \bibnamefont {Whangbo}},\ }\href
  {\doibase 10.1039/c2cc32220j} {\bibfield  {journal} {\bibinfo  {journal}
  {Chemical Communications}\ }\textbf {\bibinfo {volume} {48}},\ \bibinfo
  {pages} {5733} (\bibinfo {year} {2012})}\BibitemShut {NoStop}%
\bibitem [{\citenamefont {Chen}\ \emph {et~al.}(2015)\citenamefont {Chen},
  \citenamefont {Liu},\ and\ \citenamefont {Huang}}]{Chen2015}%
  \BibitemOpen
  \bibfield  {author} {\bibinfo {author} {\bibfnamefont {X.}~\bibnamefont
  {Chen}}, \bibinfo {author} {\bibfnamefont {L.}~\bibnamefont {Liu}}, \ and\
  \bibinfo {author} {\bibfnamefont {F.}~\bibnamefont {Huang}},\ }\href
  {\doibase 10.1039/c4cs00330f} {\bibfield  {journal} {\bibinfo  {journal}
  {Chemical Society reviews}\ }\textbf {\bibinfo {volume} {44}},\ \bibinfo
  {pages} {1861} (\bibinfo {year} {2015})}\BibitemShut {NoStop}%
\bibitem [{\citenamefont {Wang}\ \emph {et~al.}(2012)\citenamefont {Wang},
  \citenamefont {Ling}, \citenamefont {Wang}, \citenamefont {Yang},
  \citenamefont {Wang}, \citenamefont {Zhang},\ and\ \citenamefont
  {Li}}]{C2EE03158B}%
  \BibitemOpen
  \bibfield  {author} {\bibinfo {author} {\bibfnamefont {G.}~\bibnamefont
  {Wang}}, \bibinfo {author} {\bibfnamefont {Y.}~\bibnamefont {Ling}}, \bibinfo
  {author} {\bibfnamefont {H.}~\bibnamefont {Wang}}, \bibinfo {author}
  {\bibfnamefont {X.}~\bibnamefont {Yang}}, \bibinfo {author} {\bibfnamefont
  {C.}~\bibnamefont {Wang}}, \bibinfo {author} {\bibfnamefont {J.~Z.}\
  \bibnamefont {Zhang}}, \ and\ \bibinfo {author} {\bibfnamefont
  {Y.}~\bibnamefont {Li}},\ }\href {\doibase 10.1039/C2EE03158B} {\bibfield
  {journal} {\bibinfo  {journal} {Energy Environ. Sci.}\ }\textbf {\bibinfo
  {volume} {5}},\ \bibinfo {pages} {6180} (\bibinfo {year} {2012})}\BibitemShut
  {NoStop}%
\bibitem [{Asc(2012)}]{Aschauer2012}%
  \BibitemOpen
  \href {\doibase 10.1039/c2cp42288c} {\bibfield  {journal} {\bibinfo
  {journal} {Physical Chemistry Chemical Physics}\ }\textbf {\bibinfo {volume}
  {14}},\ \bibinfo {pages} {16595} (\bibinfo {year} {2012})}\BibitemShut
  {NoStop}%
\bibitem [{\citenamefont {Raghunath}\ \emph {et~al.}(2013)\citenamefont
  {Raghunath}, \citenamefont {Huang},\ and\ \citenamefont {Lin}}]{Raghunath}%
  \BibitemOpen
  \bibfield  {author} {\bibinfo {author} {\bibfnamefont {P.}~\bibnamefont
  {Raghunath}}, \bibinfo {author} {\bibfnamefont {W.~F.}\ \bibnamefont
  {Huang}}, \ and\ \bibinfo {author} {\bibfnamefont {M.~C.}\ \bibnamefont
  {Lin}},\ }\href {http://dx.doi.org/10.1063/1.4799800} {\bibfield  {journal}
  {\bibinfo  {journal} {The Journal of Chemical Physics}\ }\textbf {\bibinfo
  {volume} {138}},\ \bibinfo {eid} {154705} (\bibinfo {year}
  {2013})}\BibitemShut {NoStop}%
\bibitem [{\citenamefont {Liu}\ \emph {et~al.}(2013)\citenamefont {Liu},
  \citenamefont {Yu}, \citenamefont {Chen}, \citenamefont {Mao},\ and\
  \citenamefont {Shen}}]{Liu2013}%
  \BibitemOpen
  \bibfield  {author} {\bibinfo {author} {\bibfnamefont {L.}~\bibnamefont
  {Liu}}, \bibinfo {author} {\bibfnamefont {P.~Y.}\ \bibnamefont {Yu}},
  \bibinfo {author} {\bibfnamefont {X.}~\bibnamefont {Chen}}, \bibinfo {author}
  {\bibfnamefont {S.~S.}\ \bibnamefont {Mao}}, \ and\ \bibinfo {author}
  {\bibfnamefont {D.~Z.}\ \bibnamefont {Shen}},\ }\href {\doibase
  10.1103/PhysRevLett.111.065505} {\bibfield  {journal} {\bibinfo  {journal}
  {Physical Review Letters}\ }\textbf {\bibinfo {volume} {111}},\ \bibinfo
  {pages} {065505} (\bibinfo {year} {2013})}\BibitemShut {NoStop}%
\bibitem [{\citenamefont {Yin}\ \emph {et~al.}(2010)\citenamefont {Yin},
  \citenamefont {Chen}, \citenamefont {Yang}, \citenamefont {Gong},
  \citenamefont {Yan},\ and\ \citenamefont {Wei}}]{Yin2010}%
  \BibitemOpen
  \bibfield  {author} {\bibinfo {author} {\bibfnamefont {W.-J.}\ \bibnamefont
  {Yin}}, \bibinfo {author} {\bibfnamefont {S.}~\bibnamefont {Chen}}, \bibinfo
  {author} {\bibfnamefont {J.-H.}\ \bibnamefont {Yang}}, \bibinfo {author}
  {\bibfnamefont {X.-G.}\ \bibnamefont {Gong}}, \bibinfo {author}
  {\bibfnamefont {Y.}~\bibnamefont {Yan}}, \ and\ \bibinfo {author}
  {\bibfnamefont {S.-H.}\ \bibnamefont {Wei}},\ }\href {\doibase
  10.1063/1.3430005} {\bibfield  {journal} {\bibinfo  {journal} {Applied
  Physics Letters}\ }\textbf {\bibinfo {volume} {96}},\ \bibinfo {pages}
  {221901} (\bibinfo {year} {2010})}\BibitemShut {NoStop}%
\bibitem [{\citenamefont {Perdew}\ \emph {et~al.}(1996)\citenamefont {Perdew},
  \citenamefont {Burke},\ and\ \citenamefont {Ernzerhof}}]{pbe1997}%
  \BibitemOpen
  \bibfield  {author} {\bibinfo {author} {\bibfnamefont {J.~P.}\ \bibnamefont
  {Perdew}}, \bibinfo {author} {\bibfnamefont {K.}~\bibnamefont {Burke}}, \
  and\ \bibinfo {author} {\bibfnamefont {M.}~\bibnamefont {Ernzerhof}},\ }\href
  {\doibase 10.1103/PhysRevLett.77.3865} {\bibfield  {journal} {\bibinfo
  {journal} {Phys. Rev. Lett.}\ }\textbf {\bibinfo {volume} {77}},\ \bibinfo
  {pages} {3865} (\bibinfo {year} {1996})}\BibitemShut {NoStop}%
\bibitem [{\citenamefont {Mortensen}\ \emph {et~al.}(2005)\citenamefont
  {Mortensen}, \citenamefont {Hansen},\ and\ \citenamefont
  {Jacobsen}}]{PhysRevB.71.035109}%
  \BibitemOpen
  \bibfield  {author} {\bibinfo {author} {\bibfnamefont {J.~J.}\ \bibnamefont
  {Mortensen}}, \bibinfo {author} {\bibfnamefont {L.~B.}\ \bibnamefont
  {Hansen}}, \ and\ \bibinfo {author} {\bibfnamefont {K.~W.}\ \bibnamefont
  {Jacobsen}},\ }\href {\doibase 10.1103/PhysRevB.71.035109} {\bibfield
  {journal} {\bibinfo  {journal} {Phys. Rev. B}\ }\textbf {\bibinfo {volume}
  {71}},\ \bibinfo {pages} {035109} (\bibinfo {year} {2005})}\BibitemShut
  {NoStop}%
\bibitem [{\citenamefont {Giannozzi}\ \emph {et~al.}(2009)\citenamefont
  {Giannozzi} \emph {et~al.}}]{giannozzi}%
  \BibitemOpen
  \bibfield  {author} {\bibinfo {author} {\bibfnamefont {P.}~\bibnamefont
  {Giannozzi}} \emph {et~al.},\ }\href {http://www.quantum-espresso.org}
  {\bibfield  {journal} {\bibinfo  {journal} {Journal of Physics: Condensed
  Matter}\ }\textbf {\bibinfo {volume} {21}},\ \bibinfo {pages} {395502}
  (\bibinfo {year} {2009})}\BibitemShut {NoStop}%
\bibitem [{\citenamefont {Lejaeghere}\ \emph {et~al.}(2016)\citenamefont
  {Lejaeghere} \emph {et~al.}}]{Lejaeghere2016}%
  \BibitemOpen
  \bibfield  {author} {\bibinfo {author} {\bibfnamefont {K.}~\bibnamefont
  {Lejaeghere}} \emph {et~al.},\ }\href
  {http://science.sciencemag.org/content/351/6280/aad3000} {\bibfield
  {journal} {\bibinfo  {journal} {Science}\ }\textbf {\bibinfo {volume}
  {351}},\ \bibinfo {pages} {1415} (\bibinfo {year} {2016})}\BibitemShut
  {NoStop}%
\bibitem [{\citenamefont {Hybertsen}\ and\ \citenamefont
  {Louie}(1986)}]{PhysRevB.34.5390}%
  \BibitemOpen
  \bibfield  {author} {\bibinfo {author} {\bibfnamefont {M.~S.}\ \bibnamefont
  {Hybertsen}}\ and\ \bibinfo {author} {\bibfnamefont {S.~G.}\ \bibnamefont
  {Louie}},\ }\href {\doibase 10.1103/PhysRevB.34.5390} {\bibfield  {journal}
  {\bibinfo  {journal} {Phys. Rev. B}\ }\textbf {\bibinfo {volume} {34}},\
  \bibinfo {pages} {5390} (\bibinfo {year} {1986})}\BibitemShut {NoStop}%
\bibitem [{\citenamefont {Deslippe}\ \emph {et~al.}(2012)\citenamefont
  {Deslippe}, \citenamefont {Samsonidze}, \citenamefont {Strubbe},
  \citenamefont {Jain}, \citenamefont {Cohen},\ and\ \citenamefont
  {Louie}}]{Deslippe20121269}%
  \BibitemOpen
  \bibfield  {author} {\bibinfo {author} {\bibfnamefont {J.}~\bibnamefont
  {Deslippe}}, \bibinfo {author} {\bibfnamefont {G.}~\bibnamefont
  {Samsonidze}}, \bibinfo {author} {\bibfnamefont {D.~A.}\ \bibnamefont
  {Strubbe}}, \bibinfo {author} {\bibfnamefont {M.}~\bibnamefont {Jain}},
  \bibinfo {author} {\bibfnamefont {M.~L.}\ \bibnamefont {Cohen}}, \ and\
  \bibinfo {author} {\bibfnamefont {S.~G.}\ \bibnamefont {Louie}},\ }\href
  {\doibase http://dx.doi.org/10.1016/j.cpc.2011.12.006} {\bibfield  {journal}
  {\bibinfo  {journal} {Computer Physics Communications}\ }\textbf {\bibinfo
  {volume} {183}},\ \bibinfo {pages} {1269 } (\bibinfo {year}
  {2012})}\BibitemShut {NoStop}%
\bibitem [{Note1()}]{Note1}%
  \BibitemOpen
  \bibinfo {note} {The definition of the average electrostatic potential
  depends on the volume of the unit cell. Therefore, strictly speaking, one
  should not allow lattice vectors to change during minimization of $F$.
  However, in our calculations we find that resulting structure that minimizes
  $F$ does not depend strongly on whether lattice vectors are allowed to change
  or not.}\BibitemShut {Stop}%
\bibitem [{\citenamefont {Nelder}\ and\ \citenamefont {Mead}(1965)}]{Nelder}%
  \BibitemOpen
  \bibfield  {author} {\bibinfo {author} {\bibfnamefont {J.~A.}\ \bibnamefont
  {Nelder}}\ and\ \bibinfo {author} {\bibfnamefont {R.}~\bibnamefont {Mead}},\
  }\href {\doibase 10.1093/comjnl/7.4.308} {\bibfield  {journal} {\bibinfo
  {journal} {The Computer Journal}\ }\textbf {\bibinfo {volume} {7}},\ \bibinfo
  {pages} {308} (\bibinfo {year} {1965})}\BibitemShut {NoStop}%
\bibitem [{\citenamefont {Li}\ and\ \citenamefont
  {Wang}(2005)}]{PhysRevB.72.125325}%
  \BibitemOpen
  \bibfield  {author} {\bibinfo {author} {\bibfnamefont {J.}~\bibnamefont
  {Li}}\ and\ \bibinfo {author} {\bibfnamefont {L.-W.}\ \bibnamefont {Wang}},\
  }\href {\doibase 10.1103/PhysRevB.72.125325} {\bibfield  {journal} {\bibinfo
  {journal} {Phys. Rev. B}\ }\textbf {\bibinfo {volume} {72}},\ \bibinfo
  {pages} {125325} (\bibinfo {year} {2005})}\BibitemShut {NoStop}%
\end{thebibliography}%

\end{document}